\documentclass[twoside,leqno,twocolumn]{article}
\usepackage{amsmath,bm}
\usepackage{amsfonts}
\usepackage{latexsym}
\usepackage{array}
\usepackage{graphicx}
\usepackage{url}
\usepackage{verbatim}
\usepackage{multirow}
\usepackage{threeparttable}
\usepackage{fixltx2e}
\usepackage[center]{caption} 

\begin{document}

\title{\Large Paradigm shifts. Part I. Collagen.\\
Confirming and complementing the work of Henry Small.}
\author{Johannes Stegmann\footnote{Member of the Ernst-Reuter-Gesellschaft der Freunde, F\"orderer und Ehemaligen  der Freien Universit\"at Berlin e.V., Berlin, Germany, johannes.stegmann@fu-berlin.de}}
\date{}
\maketitle
\begin{abstract} The paradigm shift in collagen research during the early 1970s marked by the discovery of the collagen precursor molecule procollagen was traced using co-citation analysis and title word frequency determination, confirming previous work performed by Henry Small.  
\end{abstract}

\section{Introduction} \renewcommand*{\thefootnote}{\fnsymbol{footnote}}
\indent
\indent Henry Small published 1977 the paper "A Co-Citation Model of a Scientific Specialty: A Longitudinal Study of Collagen Research" (Small, 1977). The paper presented an investigation of the "reference stability" of highly cited and co-cited references in the scientific literature on collagen published 1970-1974. Comparing two consecutive years Small found that the proportion  of the number of references common in both years to the number of all distinct references in the two years were considerably lower for the years 1971/1972 and 1972/1973 than for the two years 1970/1971 and 1973/1974. Looking at three year intervals not any common reference were found for 1971/1973 (Small, 1977). Referring to Thomas Kuhn's advice to look for "a shift in the distribution of the technical literature cited in the footnotes to research reports" (Kuhn, 1962) as a possible indicator of profound changes in a scientific (sub-) field, Small concluded from his findings that a "paradigm shift" in collagen research had occurred between 1971 and 1973 (Small, 1977). \\
\indent Small supports that view by telling the story of the discovery of procollagen, the precursor molecule of collagen (Small, 1977, pp. 145 - 146). Procollagen was detected 1971 by use of biochemical and molecular biological methods (Layman et al., 1971; Bellamy and Bornstein, 1971; Lapiere et al., 1971). Small's view of procollagen detection being a paradigm shift were supported by field experts (Small, 1977, pp. 154 - 155). \\
\indent Small worked on a collagen literature set which was obtained by clustering slices of the Science Citation Index database (SCI), each slice comprising one publication year. Prior to clustering of the year slices thresholds of citation and co-citation were applied, i.e. the clusters contained only records with highly cited and highly co-cited references. The applied citation and co-citation thresholds were 15 and 11. One of each year's clusters was the collagen cluster. \\
\indent It is not possible for me to download and process whole year slices from the SCI (now SCISEARCH) or the Web of Science (WoS). So, in order to repeat Henry Small's work I had to perform a keyword-based search for the topic "collagen". In WoS I found less than 4000 papers published 1969 - 1975 using collagen-specific keywords. Because "collagen" is a biomedical topic I also searched the MEDLINE subset of PUBMED, using the Medical Subject Heading (MESH\footnote[2]{MESH = Medical Subject Headings is MEDLINE's hierarchichal thesaurus.}) "collagen" as search term and retrieved more than 6000 records in 1969 - 1975. However, PUBMED/MEDLINE records lack the cited references.  \\
\indent MEDLINE is now also a part of WoS, WoS-M. WoS' implementation of MEDLINE contains MESH and its specific retrieval tools, e.g. the "exploding" feature which allows to retrieve in one step the literature on a more general term together with the literature on all more specific terms hierarchically positioned "under" the more general term (this tool is automatically invoked when searching PUBMED/MEDLINE). Many, if not most of the WoS-M records have WoS counterparts. But downloaded WoS-M records do neither have a cited-references field (field CR in WoS records) nor WoS-specific record numbers (field UT in WoS records). Fortunately, Daniele Rotolo and Loet Leydesdorff have shown a way to overcome these difficulties (Rotolo and Leydesdorff, 2014; see also the Methods section below). Thus, it is possible to conduct a MEDLINE seach in WoS-M and assign to each retrieved WoS-M record the UT of  its corresponding WoS record (if present). Then, the WoS counterparts of the WoS-M records can be retrieved (by searches for their UT's) and downloaded. Performing the MEDLINE search makes sure (by using the MESH vocabulary) that all relevant records are retrieved\footnote[3]{in fact, the search in WoS-M yielded exactly the same number of records as in PUBMED/MEDLINE, see Results.}, and the subsequent retrieval and download of the correponding WoS records make the cited references of most of the MEDLINE records available for analysis. \\
\indent So, in contrast to Small's method I retrieved the collagen-specific literature set on the basis of an a priori keyword search, and subsequently analysed the set's reference stabilities from year to year applying citation and co-citation thresholds a posteriori. \\
\indent My "complement" of Small's work is simply an analysis of title words with respect to their new occurrence or increased frequencies comparing two year pairs. \\
\indent I will show (i) that the reference stabilities of the keyword-derived collagen literature set are similar to those found by Small (1977), and (ii) that the "paradigm shift" concluded from the changes in the reference stabilities can also be traced on the basis of word frequencies.

\section{Methods} \renewcommand*{\thefootnote}{\fnsymbol{footnote}}
\subsection{Online Retrieval and Download} \indent
\indent The search for the literature on {\em collagen} of the publication years 1969 - 1975 was performed on November 18, 2014, in the MEDLINE part of the Web of Science\footnote[4]{www.webofknowledge.com} (WoS-M). Because the WoS-M records do neither contain a CR (cited references) nor an UT (WoS identifier) field (see Introduction) the routine developed by Rotolo and Leydesdorff (2014) was applied. The Rotolo-Leydesdorff script (link given in the appendix of Rotolo and Leydesdorff, 2014) collects the UT field content, i.e. the unique number of document (hidden in the html code of each WoS-M record, provided it has a WoS counterpart) "on the fly" and writes it into a file. Than, this file containing the UTs of the WoS equilvalents of WoS-M records can be used to retrieve those WoS equivalents (Rotolo and Leydesdorff, 2014). By use of the Rotolo-Leydesdorff script I retrieved the WoS equivalents of the WoS-M collagen papers. These WoS counterparts were downloaded and subjected to analysis. \\
\indent The Rotolo-Leydesdorff routine is an {\em R-script}, i.e. one must have installed the software package R (R Core Team, 2013). \\
\indent The collagen-specific papers of the freely available database PUBMED\footnote[5]{www.ncbi.nlm.nih.gov/pubmed/} (more exactly its MEDLINE\footnote[6]{when indexed with MESH terms PUBMED records become MEDLINE records.} subset) were also retrieved and downloaded (July 18, 2014).

\subsection{Reference Stability Index} \indent
\indent The reference stability index (RSI) was calculated exactly as described by Small (who calls it "stability index (SI)")
(Small, 1977). Comparing the unique cited references in the records of publication year a with those of year b, RSI is calculated by dividing the number of shared references by the number of the sum of the unique references of both years: 
\begin{displaymath}
RSI = \frac{R_{y_{a}} \cap R_{y_{b}}}{R_{y_{a}} \cup R_{y_{b}}}.
\end{displaymath}
where $R_{y_{a}}$, $R_{y_{b}}$ are the sets of cited references in year a and year b. RSI ranges from 0 (no shared references) to 1 (shared references only). Prior to RSI calculation a citation and a co-citation threshold were applied.

\subsection{Text Analysis} \indent
\indent Document frequencies of title words - i.e. the number of records with at least one instance of the word in their titles - were determined for each publication year. Stop words\footnote[7]{Stop word list for English texts, downloaded on April 19, 2007 from ftp://ftp.cs.cornell.edu/pub/smart/english.stop.} were excluded from the analysis. Because the WoS set and the PUBMED/MEDLINE set of collagen-specific papers are of different size, text analysis was performed on both sets. In this paper, only the frequencies of the word {\em procollagen} are presented.

\subsection{Programming} \indent
\indent Extraction of record field contents, data analysis and visualisation were done using homemade programs and scripts for perl (version 5.14.2) and the software package R version 2.14.1 (R Core Team, 2013). All operations were performed on a commercial PC run under Ubuntu version 12.04 LTS.

\section{Results and Discussion}
\subsection{Retrieval} \indent
\indent Using the search string {\em MH:exp\footnote[8]{turns the "exploding" feature on, see Introduction.}=collagen} 6570  records were found in the MEDLINE part of WoS (WoS-M) for the publication years 1969 - 1975. Of the 6570 collagen-specific WoS-M records 4820 (73\%) were identified by the Rotolo-Leydesdorff routine to have a WoS counterpart. These 4820 WoS equivalents  were downloaded and subjected to reference stability analysis and title word frequency determination. The search in PUBMED using the phrase {\em "Collagen"[Mesh]\footnote[9]{automatic "exploding", see Introduction} AND   ("1969/01/01"[PDAT] : "1975/12/31"[PDAT])} retrieved also 6570 records which were downloaded and subjected to title word frequency analysis. \\
\indent The distribution of these papers to their publication years is shown in Table 1. Also shown are the number of distinct references cited by the (WoS) papers of each year. The years 1969 and 1975 were included in the analysis simply to gain some more data concerning the reference stabilities of the years surrounding the years 1971 - 1973 supposed by Small to be the core years of the transition in collagen research (Small, 1977). The data presented in Table 1 show that both, papers and references, increase by a factor of about 1.6 from 1969 to 1974 or 1975. This is clearly more than the (paper) growth of the whole WoS and PUBMED which increase from 1969 to 1974/1975 by a factor of about 1.3 to 1.4 and 1.2, respectively (data not shown), perhaps indicating a major impetus given to collagen research in the early 1970s. 

\begin{table*}[htpb]\small
\caption{Collagen research 1969 - 1975}
\centering
\begin{tabular}{cccc}
\noalign{\smallskip}
\hline
\noalign{\smallskip}
Publication years & \multicolumn{2}{c}{No. of papers} & No. of distinct \\ 
                  &  WoS  & PUBMED   & cited references (WoS only) \\
\noalign{\smallskip} 
\hline
\noalign{\smallskip}
1969 & 499 & 710 & 9700  \\
1970 & 542 & 819 & 10000  \\
1971 & 628 & 865 & 11422  \\
1972 & 757 & 1040 & 13224  \\
1973 & 744 & 999 & 14301  \\
1974 & 850 & 1113 & 15931  \\
1975 & 800 & 1024 & 15704 \\
1969 - 1975 & 4820 & 6570 & 66610 \\
\noalign{\smallskip}
\hline
\end{tabular}
\end{table*}

\subsection{Reference Stability} \indent
\indent According to Small, highly cited papers indicate relevant scientific concepts and methods agreed upon by the scientific community whereas co-citation measures the degree of association between concepts (Small, 1977, p. 141). Thus, only "highly" cited/co-cited references (I call them "core" references) were included in the reference stability analysis. However, choosing too narrow limits may result in not any core reference, and choosing too wide limits may increase the "noise", i.e. may include many references not really being concept indicator papers. So I tested several citation/co-citation thresholds including the 15/11 applied by Small (Small, 1977). Table 2 lists for each year the number of core references resulting from application of various threshold values. Threshold variations result in different number of core references for the same year (Table 2). Table 2 lists also the number of core references common to two successive publication years and the corresponding reference stability index (RSI). As described in Methods, RSI is the quotient of the number of references shared by the two compared years and the number of unique references of both years. For example, applying 10/8 as citation/co-citation threshold 18 core references were found in 1970 and 16 in 1971. 7 references are identical in both years. Hence, 27 references are unique, and RSI = 0.26 (see Table 2 fourth row, column 4 - 6). Table 3 shows numbers of shared core references and resulting RSIs for three year intervals. 

\begin{table*}[htpb]
\tiny
\caption{Reference stability of core references in collagen research 1969 - 1975: two year intervals.}
\centering
\begin{threeparttable}
\begin{tabular}{cccccccccccccc}
\hline
\noalign{\smallskip}
Citation/Co-citation  & \multicolumn{13}{c}{Number of core and shared core references and reference stability indexes of two year intervals}\\
        thresholds & \\
 & 1969 & sh\tnote{1}/RSI\tnote{2} & 1970 & sh/RSI & 1971 & sh/RSI & 1972 & sh/RSI & 1973 & sh/RSI & 1974 & sh/RSI & 1975 \\

\hline
\noalign{\smallskip}
15/11 & 2 & 2/1.0 & 2 & 2/0.50 & 4 & 4/0.31 & 13 & 6/0.30 & 13 & 11/0.50 & 20 & 3/0.11 & 11 \\
15/8 & 7 & 2/0.20 & 5 & 2/0.22 & 6 & 4/0.24 & 15 & 8/0.30 & 20 & 15/0.45 & 28 & 9/0.23 & 20 \\
11/9 & 13 & 6/0.35 & 10 & 6/0.38 & 12 & 6/0.22 & 21 & 12/0.32 & 28 & 18/0.35 & 41 & 10/0.17 & 28 \\
10/8 & 19 & 6/0.19 & 18 & 7/0.26 & 16 & 10/0.28 & 30 & 16/0.27 & 45 & 28/0.38 & 57 & 20/0.25 & 44 \\
10/5 & 40 & 19/0.33 & 37 & 21/0.38 & 39 & 25/0.28 & 74 & 33/0.30 & 70 & 47/0.40 & 95 & 37/0.26 & 82 \\
\noalign{\smallskip}
\hline
\end{tabular}
\begin{tablenotes}
\footnotesize
\item[1] sh: number of core references shared by both years.
\item[2] RSI: reference stability index (see Methods).
\end{tablenotes}
\end{threeparttable}
\end{table*}

\begin{table*}[htpb]\small
\caption{Reference stability of core references in collagen research 1969 - 1975: three year intervals.}
\centering
\begin{threeparttable}
\begin{tabular}{cccccc}
\hline
\noalign{\smallskip}
Citation/Co-citation  & \multicolumn{5}{c}{Number of shared core references and RSIs of three year intervals\tnote{1,2}}\\
          thresholds & \\
 &  1969/1971 & 1970/1972 & 1971/1973 & 1972/1974 & 1973/1975 \\
\noalign{\smallskip}
\hline
\noalign{\smallskip}
15/11  & 2/0.50 & 2/0.15 & 2/0.13 & 7/0.27 & 0/0.00 \\
15/8  & 2/0.18 & 2/0.11 & 2/0.08 & 7/0.19 & 5/0.14 \\
11/9  &  5/0.25 & 4/0.15 & 3/0.08 & 11/0.22 & 9/0.19 \\
10/8  & 8/0.30 & 6/0.14 & 4/0.07 & 13/0.18 & 15/0.20 \\
10/5  & 20/0.34 & 16/0.17 & 17/0.18 & 31/0.22 & 29/0.24 \\
\noalign{\smallskip}
\hline
\end{tabular}
\begin{tablenotes}
\footnotesize
\item[1] Values are denoted in the sequence sh/RSI (see Table 2).
\item[2] For numbers of year-specific core references see Table 2.
\end{tablenotes}
\end{threeparttable}
\end{table*}

\indent From the data presented in Table 2 and Table 3 it is evident that the results yielded from the 15/11 and 15/9 
citation/co-citation thresholds applied a posteriori apparently fit best with Small's outcome of a priori application of the 15/11 thresholds (Small, 1974, p. 144). This holds true with respect to the number of shared core references as well as RSI values of two year and three year intervals (see row two in Table 2 and Table 3). However, the RSIs of the two year intervals are certainly not sign of a trend: although the values derived from the 15/11 restriction decrease from 1969/1970 (Table 2) or 1970/1971 (Small, 1977, p. 144) onward until 1972/1973, followed by an increase in 1973/1974 (Table 2, row 1; Small, 1977, p. 144), they decrease in 1974/1975 for each of the chosen thresholds as compared with the RSI of the preceding two year interval (see Table 2). In addition, looking at the RSI values derived from less restrictive citation/co-citation constraints we see increasing RSIs (except RSI for 1974/1975) under the 15/8 condition or variantly de- and increasing RSI values under the remaining thresholds (Table 2). \\
\indent Looking at the three year intervals (Table 3), however, we see each value series decreasing to its lowest RSI value in 1971/1973 (under the 10/5 restriction the lowest RSI is 0.17 for 1970/1972 followed by almost identical 0.18 for 1971/1973; see Table 3, last row). RSI values for the subsequent three year interval 1972/1974 are again all higher than the preceding ones. For the successive three year interval 1973/1975 we see an increase under low restrictions (10/8, 10/5) and a decrease under the other more stringent restrictions. \\
\indent In Small's paper the RSI of the 1971/1973 interval is zero. This is confirmed by the data presented in Table 3 where the RSI values of the same interval are the lowest ones and - for three citation/-co-citation thresholds - are with 0.08 and 0.07 near to zero. Small's interpretation of these data is that 1971/1973 was the time span during which a "conceptual shift" in collagen research had occurred (Small, 1977, p. 144). However, we must add a "caveat" to this interpretation. We simply do not have enough data for such a conclusion. One should analyse many more three year intervals to get an impression of what is possible with respect to reference stabilities. Possibly, we will not be able to statistically secure the lowest RSI in a time series as a reliable indicator of a "paradigm shift" in a specialty. On the other hand, the here applied citation/co-citation technique developed by Henry Small may be regarded as a kind of a "pocket lamp" to detect interesting phenomena in the cited references jungle of scientific publications. Having detected an interesting situation one must have a closer look at the phenomenon using also other methods, e.g. gathering experts' opinions (as Small did). \\
\indent The RSI of the three year interval 1973/1975 of the most stringent 15/11 threshold is also zero (see Table 3). A reason for that might be a possible diversification of collagen research following the "paradigm shift" of procollagen detection, thus relaxing previous tight connections within the field. Lowering the thresholds results in RSIs comparable to those of the antecedent three year interval (see Table 3), indicating a more "normal" reference stability. \\
\indent In summary, we can say both methods - Small's method of clustering whole year slices of the database under the restriction of a a priori citation/co-citation threshold followed by selection of (a) appropriate theme cluster(s) and the method used in this paper of applying thresholds a posteriori to a literature set retrieved by a keyword search - lead to similar results and are able to identify a major breakthrough in collagen research at the beginning 1970s.

\subsection{Text Analysis} \indent

\begin{figure}[htpb]
\centerline{
\includegraphics[height=8.0cm]{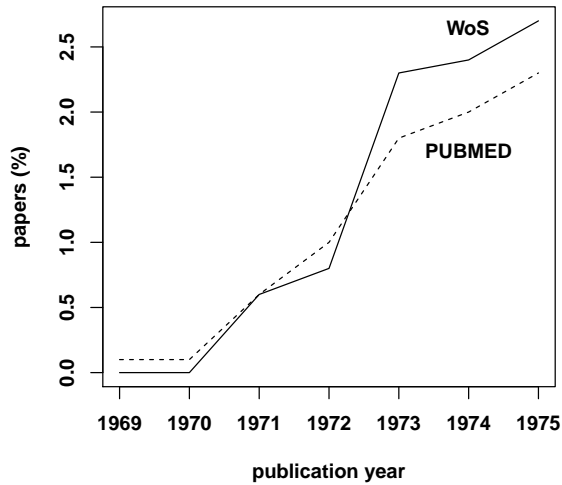}}
\caption{Title frequencies of {\em procollagen} in collagen research papers 1969 - 1975.}
\end{figure}

As described in Methods, the frequencies of title words were determined. More exactly, I looked for the occurrences of the technical term "procollagen". Figure 1 shows for each year the relative title frequency of the word {\em procollagen} in the WoS and PUBMED literature sets. In 1969 and 1979 there are no titles with {\em procollagen} in the WoS set, whereas in the PUBMED set a very low percentage of the papers has titles with the word {\em procollagen} (0.1\%, representing one paper each). From 1971 onward, however, there is a strong increase of the occurrence of {\em procollagen} as title word in both sets, ranging from 0.6\% (1971) to more than 2\% in the later years (see Figure 1). \\
\indent Reliable data of the portion of abstracts containing the word {\em procollagen} could not be obtained because most records lack an abstract. Only four records of the WoS set contain abstracts; none of them contains the word {\em procollagen}. The PUBMED set contains between 4\% and 6.8\% abstracts in 1969 - 1974; some of them mention {\em procollagen}. In 1975, 60\% of the records contain abstracts; about 4\% of them mention {\em procollagen} (not shown). \\ 
\indent The present study focuses on the word {\em procollagen} because it is the "label" of the paradigm shift in collagen research in the early 1970s. More in-depth study perhaps would find other concept-bearing words indicating a transition in research. Yet from the rudimentary analysis shown in Figure 1 we can conclude that it might be worthwhile simply to count words in order to detect changes over time. But, similar to the analysis of reference stability, the application of frequency thresholds in word analysis is certainly necessary due to the high number of words. Some of the year's set of titles of the present study comprise more than 2000 distinct words (not shown), and an entire set of abstracts surely has many more unique words. The threshold could be such that only words (or multi-word phrases) occurring in a minimal number of papers are included in the study. Co-word analysis is also possible; here, frequency thresholds and similarity thresholds (e.g. cosine values above a certain value) could be applied. Careful screening of the resulting word lists is necessary, and expert opinions are indispensable, especially when the most recent scientific literature is screened do detect beginnings of still unknown transitions. 

\section{Conclusion} \indent
\indent The study presented here has shown that the paradigm shift in collagen research during the early 1970s found by Henry Small on the basis of a pre-clustered literature set can also be traced using literature sets retrieved by an appropriate keyword search. In addition, the study shows that the shift is also indicated by the appearance of the word {\em procollagen} and its presence in a growing number of paper titles during the investigated time period.

\end{document}